\documentclass{emulateapj}
\usepackage{textcomp}
\usepackage{mathtext}
\usepackage{amssymb}

\newcommand{\fermi}{\textit{Fermi}}
\newcommand{\gr}{$\gamma$-ray}

\begin{document}

\title{Fermi Observation of the transitional pulsar binary XSS J12270$-$4859}

\author{Yi Xing and Zhongxiang Wang}

\affil{Key Laboratory for Research in Galaxies and Cosmology,
Shanghai Astronomical Observatory,\\ 
Chinese Academy of Sciences, 80 Nandan Road, Shanghai 200030, China}

\begin{abstract}

Because of the disappearance of its accretion disk 
since the time period around 2012 November--December, 
XSS J12270$-$4859 has recently been identified as, 
in addition to PSR J1023$+$0038, another transitional millisecond pulsar 
binary. We have carried out detailed analysis of 
the \textit{Fermi} Large Area Telescope data for the binary.
While both spectra before and after the disk-disappearance transition
are well described by an exponentially cut-off power law, 
typical for pulsars' emission
in the \textit{Fermi}'s 0.2--300\,GeV, a factor of 2 flux decrease
related to the transition is detected. A weak orbital modulation is seen,
but only detectable in the after-transition data, same to that found at X-rays.
In the long-term light curve of the source before the transition,
a factor of 3 flux variations are seen.
Comparing to the properties of J1023+0038, we disucss the implications
from these results.  We suggest that since the modulation is aligned 
with that at X-rays in orbital phase, it
possibly arises due to the occultation of the $\gamma$-ray emitting 
region by the companion.  The origin of 
the variations in the long-term light curve is not clear,
because the source field also contains unidentified 
radio or X-ray sources and
their contamination can not be excluded. Multi-wavelength observations
of the source field will help identify the origin of the variations by
detecting any related flux changes from the in-field sources.

\end{abstract}

\keywords{binaries: close --- stars: individual (XSS J12270$-$4859) --- stars: low-mass --- stars: neutron}

\section{INTRODUCTION}

It is clear now that during the evolution from low-mass X-ray 
binaries (LMXBs) to millisecond pulsars (MSPs; e.g., \citealt{bv91}), 
a neutron star in such a system 
can switch between the states of being accretion-powered and rotation-powered,
which is well illustrated from observational studies of 
the MSP binary J1824$-$2452I in the globular cluster M28 \citep{pap+13}.
Before, indirect evidence has strongly suggested that several accreting MSPs in
transient LMXBs probably switch to be rotation powered during their 
quiescent states (\citealt{bur+03}; \citealt{wan+13} and references therein).
The discovery of the radio MSP binary J1023$+$0038 \citep{arc+09} has further 
shown interesting features for probably the end phase of the evolution:
such a binary can switch between the states of having an accretion disk and 
being disk-free, while the pulsar in the binary is still active but not 
observable during the former state \citep{sta+14,cot+14}. PSR J1023$+$0038 
is also a prototypical,
so-called `redback' system \citep{rob13}. Different from those `black widow'
MSP binaries (e.g., B1957$+$20; \citealt{fst88}) that contain very low-mass
($\sim$0.02\,$M_{\sun}$) companions, redback systems have relatively massive
secondaries ($\sim$0.1--0.6\,$M_{\sun}$). Irradiation 
and/or evaporation of the companion by emission and pulsar wind, 
respectively, from the central neutron star probably play an important 
role in forming black widow and redback systems \citep{che+13,ben+14}.

The second transitional MSP binary, XSS J12270$-$4859, was recently identified,
as the accretion disk in the system was found to have disappeared since 2012 
Nov.--Dec. \citep{bas+14,mar+14}.  Discovered by the Rossi X-ray Timing 
Explorer \citep{sr04} and initially thought as a cataclysmic 
variable \citep{mas+06}, the binary was considered as a peculiar LMXB 
that is associated with the \gr\ source 2FGL J1227.7$-$4853
(\citealt{hil+11,mar+13} and references therein). After its transition
to be disk-free, multiple energy observations have revealed its
full nature. The detection of a 1.69 ms radio pulsar was
reported \citep{roy+14,ray+14}. Optical observations have found that its 
orbital period is 6.9\,hr \citep{bas+14,mar+14}, and the 
companion is a mid G-type, under-massive donor star with mass of 
0.06--0.12\,$M_{\sun}$ \citep{mar+14}, establishing it
as another redback system. 
At X-ray energies, the binary is an order of magnitude fainter 
than before, with orbital modulation present \citep{bog+14}.
Comparing to that seen in J1023+0038 in the disk-free 
state \citep{arc+10,bog+11}, the X-ray emission likely has an 
intrabinary-shock origin \citep{bog+14}, due to the interaction between 
the outflow from the companion and the pulsar wind.
At \gr\ energies, a possible flux decrease was also reported in
an astronomer's telegram \citep{tam+13}. 
These properties are highly comparable to that of J1023+0038, particularly
since the latter system just underwent a transition to the active accretion
state in 2013 June and extensive observational studies have been conducted
\citep{sta+14,pat+14,tak+14,cot+14}.

The source field of 2FGL J1227.7$-$4853 was found to contain three 
radio sources \citep{hil+11}. The brightest one J122806$-$485218 has extended 
radio emission, although its radio properties 
likely do not support the classification
of it being a blazer \citep{hil+11}, the class for which is the major 
population ($\sim$80\%) detected by \textit{Fermi} in its
Large Area Telescope (LAT)
second source catalog \citep{nol+12}. 
X-ray observations did not reveal any physical connection 
between XSS J12270$-$4859 and J122806$-$485216, but found a possible, 
previously uncatalogued galaxy cluster that is located approximately 1\arcmin\ 
away from the former \citep{bog+14}.  
Previously, \citet{hil+11} and \citet{mar+13} have analyzed approximately
2-yr and 4-yr \textit{Fermi} data for 2FGL J1227.7$-$4853, respectively, and
mainly studied its \gr\ spectral properties.

Given above and the recent identification of the transitional MSP binary 
nature for XSS~J12270$-$4859, detailed analysis of the \textit{Fermi} data 
is needed, which possibly helps our understanding of the physical processes 
in this system. We have carried 
out the analysis to identify the properties of the associated
\gr\ source, study its long-term flux variability,
and search for modulation signals.
In this paper, we report the results from the analysis. 
We describe the \textit{Fermi} observation data in \S~\ref{sec:obs}, 
and present the data analysis and results in \S~\ref{sec:ana}. 
Discussion of the results is given in \S~\ref{sec:disc}.

\section{Observation}   
\label{sec:obs}

LAT is a $\gamma$-ray imaging instrument onboard \textit{Fermi}.
It makes all-sky survey 
in an energy range from 20 MeV to 300 GeV \citep{atw+09}. In our analysis,
we selected LAT events from the \textit{Fermi} Pass 7 Reprocessed (P7REP) 
database inside a $\mathrm{20^{o}\times20^{o}}$ region centered 
at the position of XSS J12270$-$4859, which is 
R.A.=186\fdg994783, Decl.=$-$48\fdg895244 (equinox J2000.0) obtained from 
the Two Micron All Sky Survey (2MASS, \citealt{cut+03}) and used 
in \citet{bog+14} for X-ray studies of orbital modulation of this source. 
We kept events during the time period from 2008-08-04 15:43:36 (UTC) to 
2014-07-10 18:16:37 (UTC) and in the energy range of 100\,MeV to 300\,GeV. 
In addition, only events with zenith angle less than 100\,deg  
and during good time intervals were selected. The former prevents 
the Earth's limb contamination, and for the latter, the quality of 
the data was not affected by the spacecraft events. 

\section{Data Analysis and Results} 
\label{sec:ana}

\subsection{Source Identification}
\label{subsec:si}

We included all sources within 16 deg in the \textit{Fermi} second 
source catalog \citep{nol+12} centered at the position of 
XSS J12270$-$4859 to make the source model. The spectral function 
forms of the sources are provided in the catalog. 
The spectral normalization parameters of the sources 
within 8 deg from XSS J12270$-$4859 were set free, and 
all other parameters of the sources were fixed at their catalog values. 
We used the spectrum model gll\_iem\_v05\_rev1.fits and the spectrum 
file iso\_source\_v05.txt for the Galactic and the extragalactic diffuse 
emission, respectively, in the source model. 
The normalizations of the diffuse components were set as free parameters.

\begin{figure}
\centering
\epsscale{1.0}
\plotone{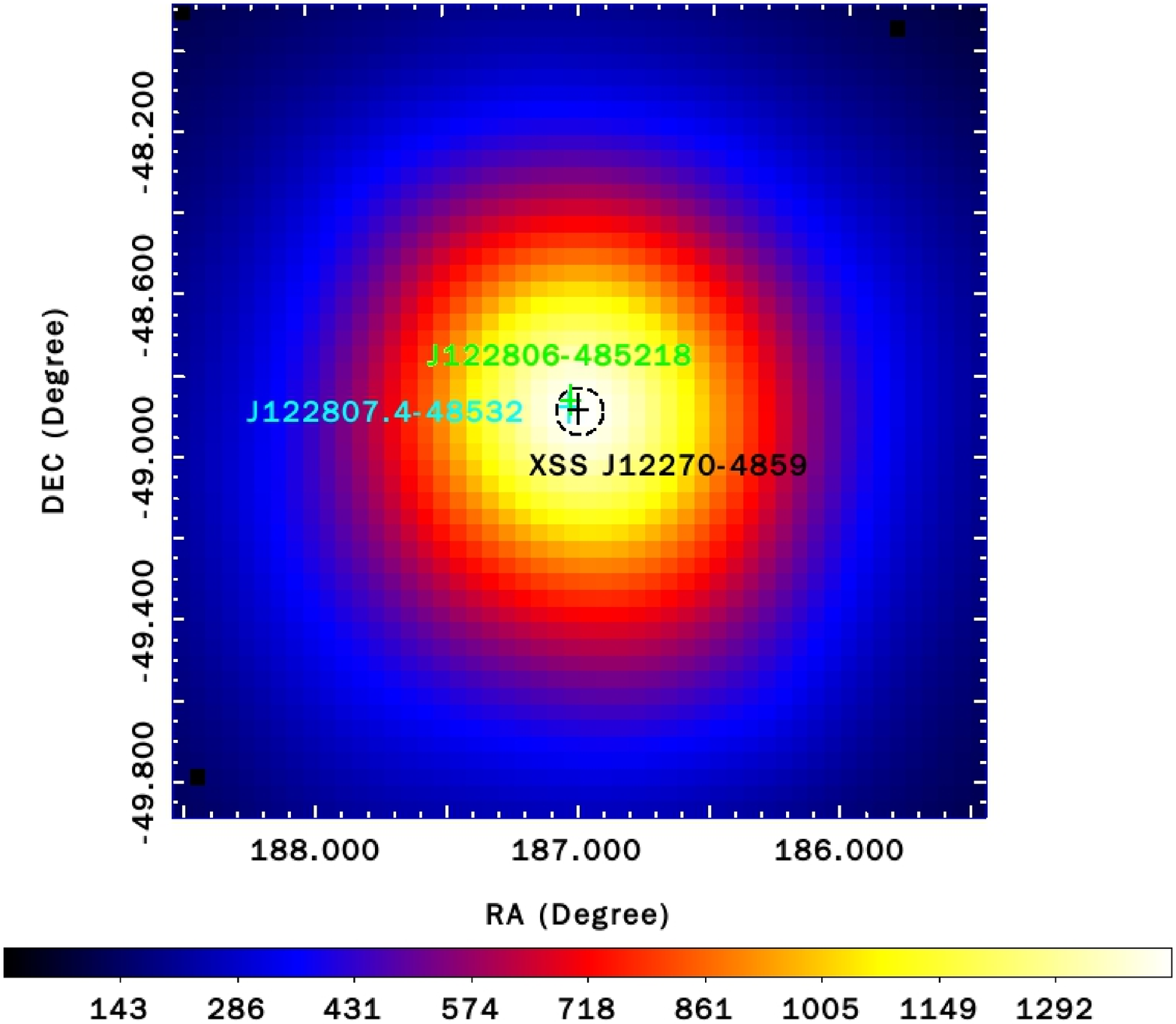}
\caption{200 MeV$-$300 GeV TS map of a $\mathrm{2^{o}\times2^{o}}$ 
region centered at the position of XSS J12270$-$4859. The image scale 
of the map is 0.04\arcdeg\ pixel$^{-1}$. All sources in the source model 
were considered and removed. 
The dark cross, green cross, light blue cross, and the dashed black circle 
mark the 2MASS position of XSS J12270$-$4859, the nearby brightest radio source, the nearby galaxy cluster, 
and the 2$\sigma$ error circle of the best-fit position obtained from 
using the \textit{Fermi} data, respectively.}
\label{fig:tsmap}
\end{figure}

Using the LAT science tools software package {\tt v9r33p0}, we performed
standard binned likelihood analysis to the LAT data in the $>$0.2\,GeV range.
Events below 200 MeV were rejected
because of the relative large uncertainties of the instrument response 
function of the LAT in the low energy range. We extracted the 
Test Statistic (TS) map of a $2\arcdeg\times 2\arcdeg$ region centered 
at the position of XSS J12270$-$4859 (Figure~\ref{fig:tsmap}), with all 
sources in the source model considered except the candidate $\gamma$-ray 
counterpart (2FGL J1227.7$-$4853) to XSS J12270$-$4859.
The TS value at a specific position, calculated from 
TS$= -2\log(L_{0}/L_{1})$ (where $L_{0}$ 
and $L_{1}$ are the maximum likelihood values for a model without and with 
an additional source respectively), is a
measurement of the fit improvement for including the source, and
is approximately the square of the detection significance 
of the source \citep{1fgl}.

As shown in Figure~\ref{fig:tsmap}, the $\gamma$-ray emission near 
the center was detected with TS$\simeq$1440, indicating $\sim$37$\sigma$ 
detection significance.  
We ran \textit{gtfindsrc} in the LAT software package to find 
the position of the $\gamma$-ray source
and obtained R.A.=186\fdg99, Decl.=$-$48\fdg90 
(equinox J2000.0), with 1$\sigma$ nominal uncertainty of 0\fdg03.
This best-fit position is consistent with
the 2MASS position of XSS J12270$-$4859 (mark by a dark cross 
in Figure~\ref{fig:tsmap}). The offset between the two positions
is only $\sim$0\fdg01. 
The nearby radio source J122806$-$485218 and possible galaxy cluster 
J122807.4$-$48532 are also within the 2$\sigma$ error circle, 
which are $\simeq$0\fdg04 and $\simeq$0\fdg03 away from the best-fit position, 
respectively. 

Including the \gr\ source in the source model 
at the 2MASS position of XSS J12270$-$4859, 
we performed standard binned likelihood analysis to the LAT data 
in $>$0.2 GeV range, with emission of 
this source modeled with a simple power law and an exponentially cutoff 
power law (characteristic of pulsars). 
For the former model, photon index $\Gamma= 2.41\pm$0.03 with a
TS$_{pl}$ value of $\sim$1446 was obtained, and for the latter,
$\Gamma= 2.11\pm$0.08 and cutoff energy $E_c=6\pm$2 GeV, with a TS$_{exp}$ 
value of $\sim$1466, were obtained.
These results are also given in Table~\ref{tab:likelihood}.
The low-energy cutoff was thus detected with 
$>$4$\sigma$ significance (estimated from $\sqrt{{\rm TS}_{cutoff}}$,
where TS$_{cutoff}\simeq {\rm TS}_{exp}-{\rm TS}_{pl}\simeq 20$).
The result favors the association of the $\gamma$-ray source with the
pulsar binary system XSS J12270$-$4859.

\subsection{Variability Analysis}
\label{subsec:lv}

\begin{figure}
\centering
\epsscale{1.0}
\plotone{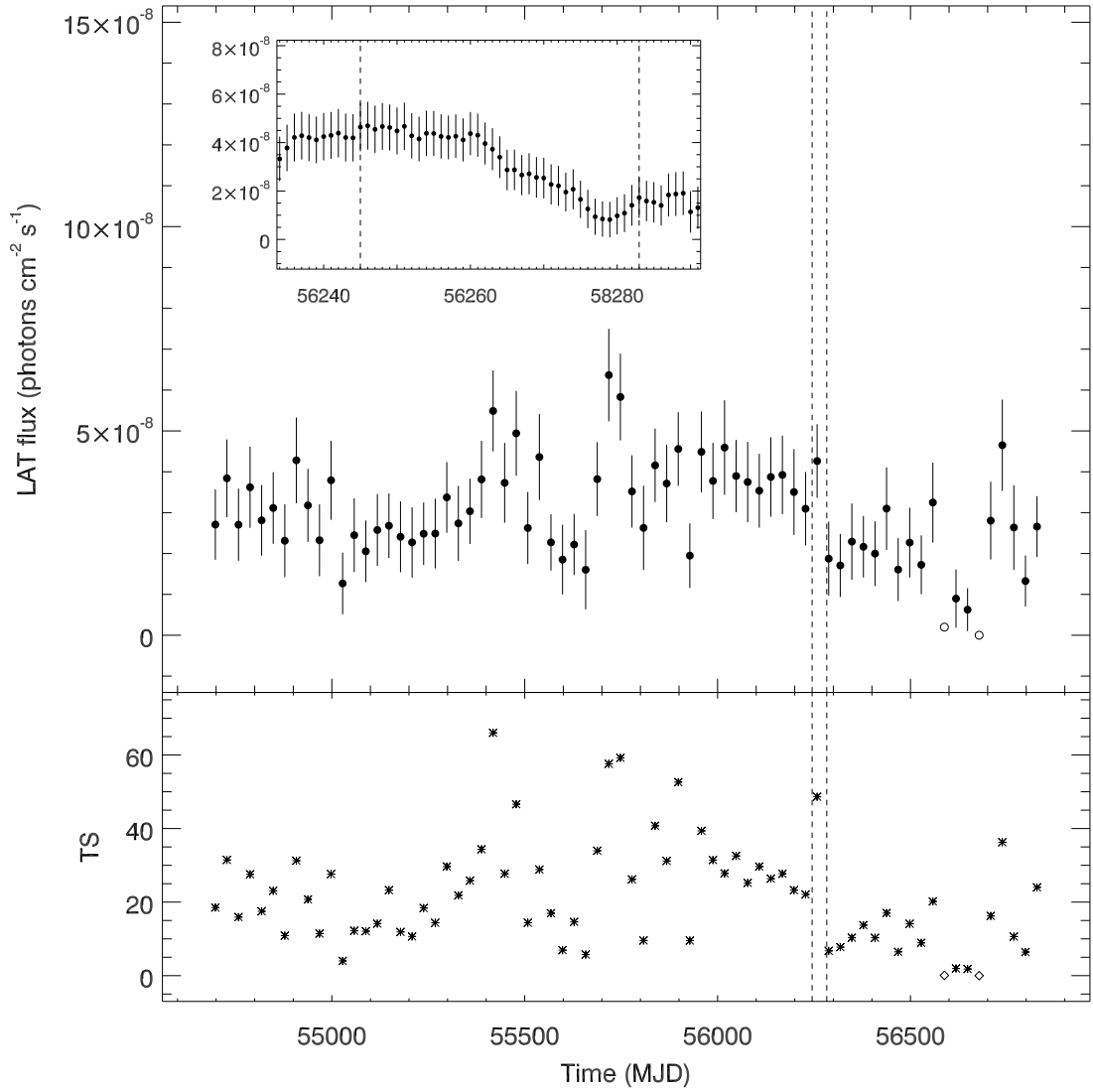}
\caption{30-day interval light curve and TS curve of 2FGL J1227.7$-$4853. 
The dashed lines mark the time period of the state transition, and
the inset panel shows the detailed flux changes during the time period.}
\label{fig:lc}
\end{figure}

We obtained the light curve of the source in $>$0.2 GeV energy range 
to search for flux variations,
particularly around the state-transition time period of XSS J12270$-$4859.
A point source with $\Gamma= 2.41$ power-law emission (\S~\ref{subsec:si})
at the 2MASS position was considered.
Significant variations before the state 
change (MJD 56245) are seen in the 30-day time interval light curve. 
In Figure~\ref{fig:lc}, the light curve as well as the TS curve
are shown. The flux has a factor of 3 variations 
(2--6$\times$10$^{-8}$\,photons\,cm$^{-2}$ s$^{-1}$), while
TS varies between 10--70.
After the state change (around MJD 56283), $\gamma$-ray emission was 
relatively stable, with the flux and TS being 
$\sim$2$\times$10$^{-8}$\,photons\,cm$^{-2}$ s$^{-1}$ and $\sim$20 
for most of the time. 
We noted that there were two unreliable data points at the time bins
of MJD 56573--56603 and 56663--56693. The obtained TS values are smaller
than 1 (marked by open symbols in Figure~\ref{fig:lc}). 
During the times, there were two target-of-opportunity observations causing
data gaps of $\sim$5 and $\sim$12 days, respectively.

To further investigate the flux variations during the transition time period,
we constructed a 30-day light curve by shifting each time interval by 1 day
forward and obtaining the flux during such a 30-day interval
(see, e.g., \citealt{tak+14}). The resulting fine smooth light curve
is shown in the inset box of Figure~\ref{fig:lc}. 
A factor of 2 flux decrease, from $\sim 4\times 10^{-8}$ to 
$\sim 2\times 10^{-8}$\,photons\,cm$^{-2}$ s$^{-1}$, is clearly visible.
Therefore 2FGL~J1227.7$-$4853 had a flux change related to the state transition.

We performed standard binned likelihood analysis to the LAT data 
in $>$0.2 GeV range for the time periods before and after 
the state change. 
When a power-law spectrum was assumed for the former data, 
we found $\Gamma= 2.42\pm0.04$ with TS$_{pl}\simeq$ 1247
and for the latter, $\Gamma= 2.42\pm0.09$ with a TS$_{pl}\simeq$ 183.
When an exponentially cutoff power law was considered, the respective
results were $\Gamma= 2.13\pm$0.08 and $E_c= 6\pm$2 GeV with 
TS$_{exp}\simeq$ 1262, and $\Gamma= 1.8\pm$0.3 and $E_c= 2\pm$1 GeV 
with TS$_{exp}\simeq$ 192.
Therefore the low-energy cutoff was detected with TS$_{cutoff}$ values 
of $\simeq$15 and $\simeq$9, which correspond to the detection significance 
of $\simeq 3.8\sigma$ and $\simeq 3\sigma$ before and after the state change, 
respectively. These results indicate that the $\gamma$-ray emission from 
the source is likely better described by a spectrum characteristic of pulsars, 
and the emission after the state transition is harder. 
For the exponentially cutoff power-law spectra, 
the $>$0.2 GeV $\gamma$-ray fluxes were 
3.2$\pm$0.2 $\times$ 10$^{-8}$ and 1.7$\pm$0.2 $\times$ 10$^{-8}$ 
photons\,cm$^{-2}$\,s$^{-1}$ during the time periods before and after the state change, respectively.
These results are summarized in Table~\ref{tab:likelihood}.

\subsection{Spectral Analysis}
\label{subsec:sa}

\begin{figure}
\centering
\epsscale{1.0}
\plottwo{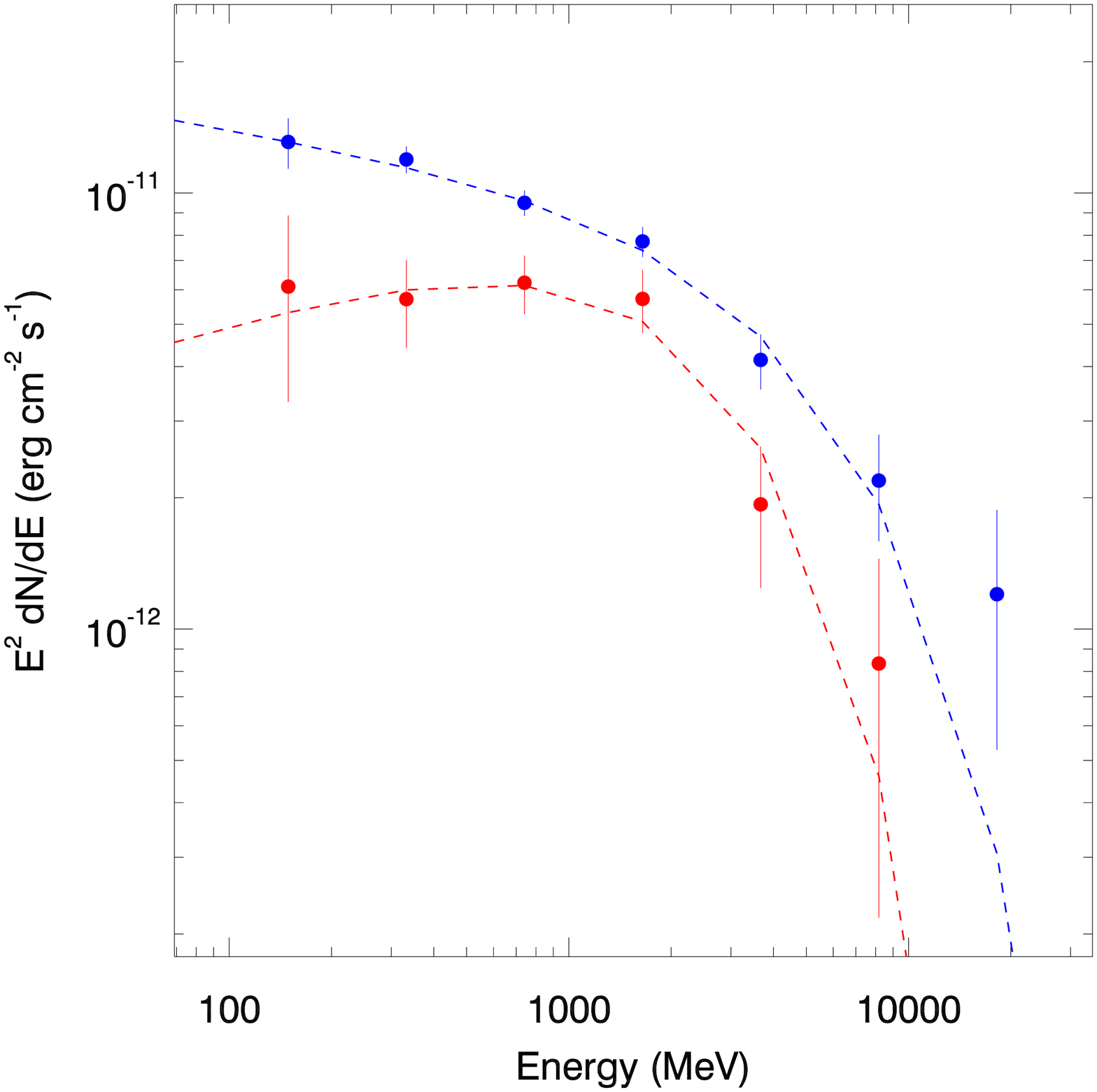}{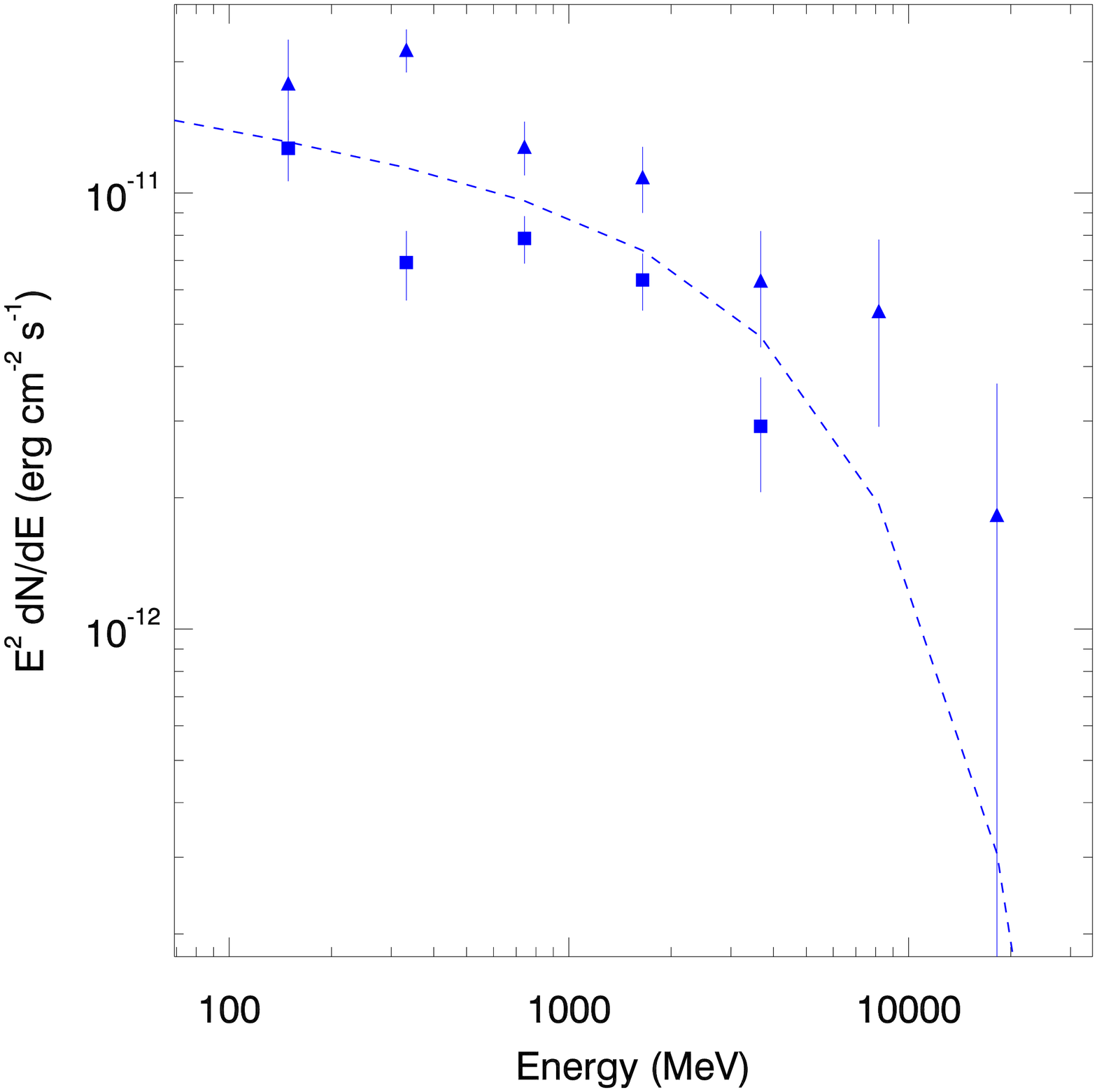}
\caption{{\it Left panel:} $\gamma$-ray spectra of 2FGL J1227.7$-$4853 
obtained before (blue data points) and after (red data points) the state 
transition. {\it Right panel:} $\gamma$-ray spectra of 2FGL J1227.7$-$4853 
obtained during the high (triangle data points) and
low (square data points) states. The exponentially cutoff power laws
obtained from maximum likelihood analysis before and after the state transition 
are shown as blue and red dashed curves.}
\label{fig:spectrum}
\end{figure}

We report our spectrum results of the \gr\ source by 
considering the emission as a point source with a power-law 
spectrum at the 2MASS position. The photon index was fixed to the value 
we obtained above using the total data (see Table~\ref{tab:likelihood}).
The spectrum was extracted by
performing maximum likelihood analysis to the LAT data 
in 10 evenly divided energy bands in logarithm from 0.1--300 GeV. 
Only spectral points with TS$\geq$4 were kept.
We extracted $\gamma$-ray spectra for 
the time periods before and after the state change, respectively. 
The obtained spectra are shown in the left panel of 
Figure~\ref{fig:spectrum}, and the energy flux values are given 
in Table~\ref{tab:spectrum-point}. 

In order to investigate the variability before the state transition, 
we also extracted $\gamma$-ray spectra from the data in the time intervals
of TS$\geq$35 (`high' state) and TS$\leq$20 (`low' state).
The obtained spectra are shown in the right panel of 
Figure~\ref{fig:spectrum}, with the energy 
flux values given in Table~\ref{tab:spectrum-point}. Comparing
the two spectra, while the source appeared brighter across 
the whole energy range in the high state, the flux at
0.33\,GeV, showing a $\simeq$5$\sigma$ difference,
had the most significant change. 

\subsection{Timing Analysis}
\label{subsec:ta}

\begin{figure}
\centering
\epsscale{1.0}
\plotone{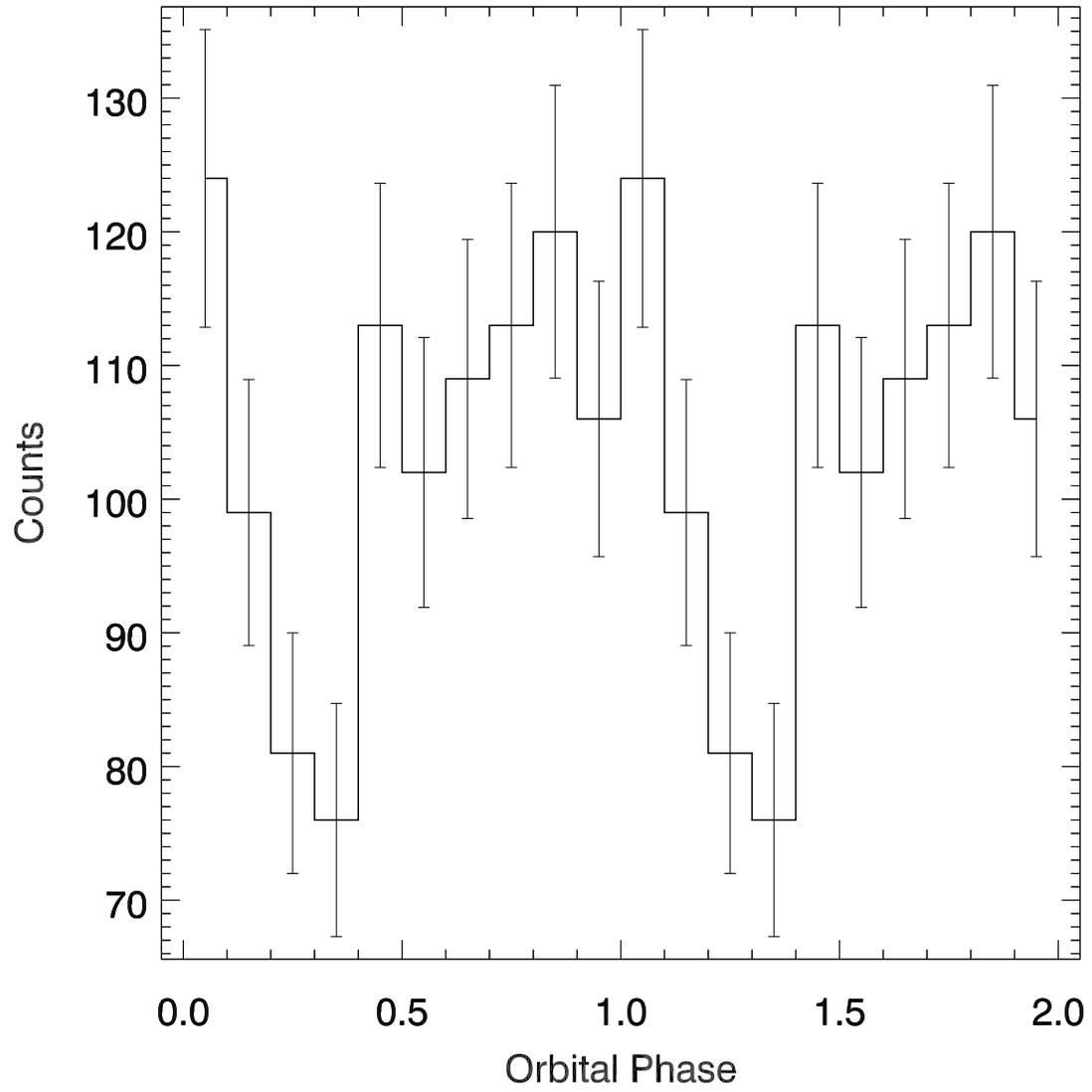}
\caption{0.3--300 GeV light curve folded at the optical orbital period,
obtained from the data after the state transition.}
\label{fig:timing}
\end{figure}

We performed timing analysis to the LAT data of XSS J12270$-$4859 to search 
for possible orbital modulations. Considering the X-ray orbital modulation 
was only detected after the state change \citep{bog+14}, we first folded 
the LAT data of the \gr\ source during the time period at the optical 
orbital frequency of 4.01850$\pm 0.00003\times 10^{-5}$\,Hz (\citealt{mar+14};
see also \citealt{bas+14}).
The 2MASS position of XSS J12270$-$4859 was used for 
the barycentric corrections to photon arrival times, and photons 
within 1\fdg2 from the position were collected. 
Different energy ranges (e.g., $>$0.2, $>$0.3, $>$0.5, $>$1, and $>$2 GeV)
were tested in folding.
We found that the highest orbital signal was revealed in the 
$>$0.3 GeV energy range. The folded light curve, which has
an H-test value of $\sim$10 (corresponding to 3$\sigma$ detection 
significance), is shown in Figure~\ref{fig:timing}. The phase 
zero is set at the ascending node of the pulsar in XSS J12270$-$4859. 
Although the significance is
not high, we note that the folded light curve is nearly 
aligned with the X-ray one given in \citet{bog+14}, which 
strengthens the modulation detection. 
The result thus provides strong evidence for the association of 
the $\gamma$-ray source with XSS J12270$-$4859.

As a test, we also folded the LAT data of XSS J12270$-$4859 before 
the state change at the optical orbital frequency.
No significant $\gamma$-ray modulation was detected (the
H-test values were $\sim$0.3).

\begin{figure}
\centering
\epsscale{1.0}
\plottwo{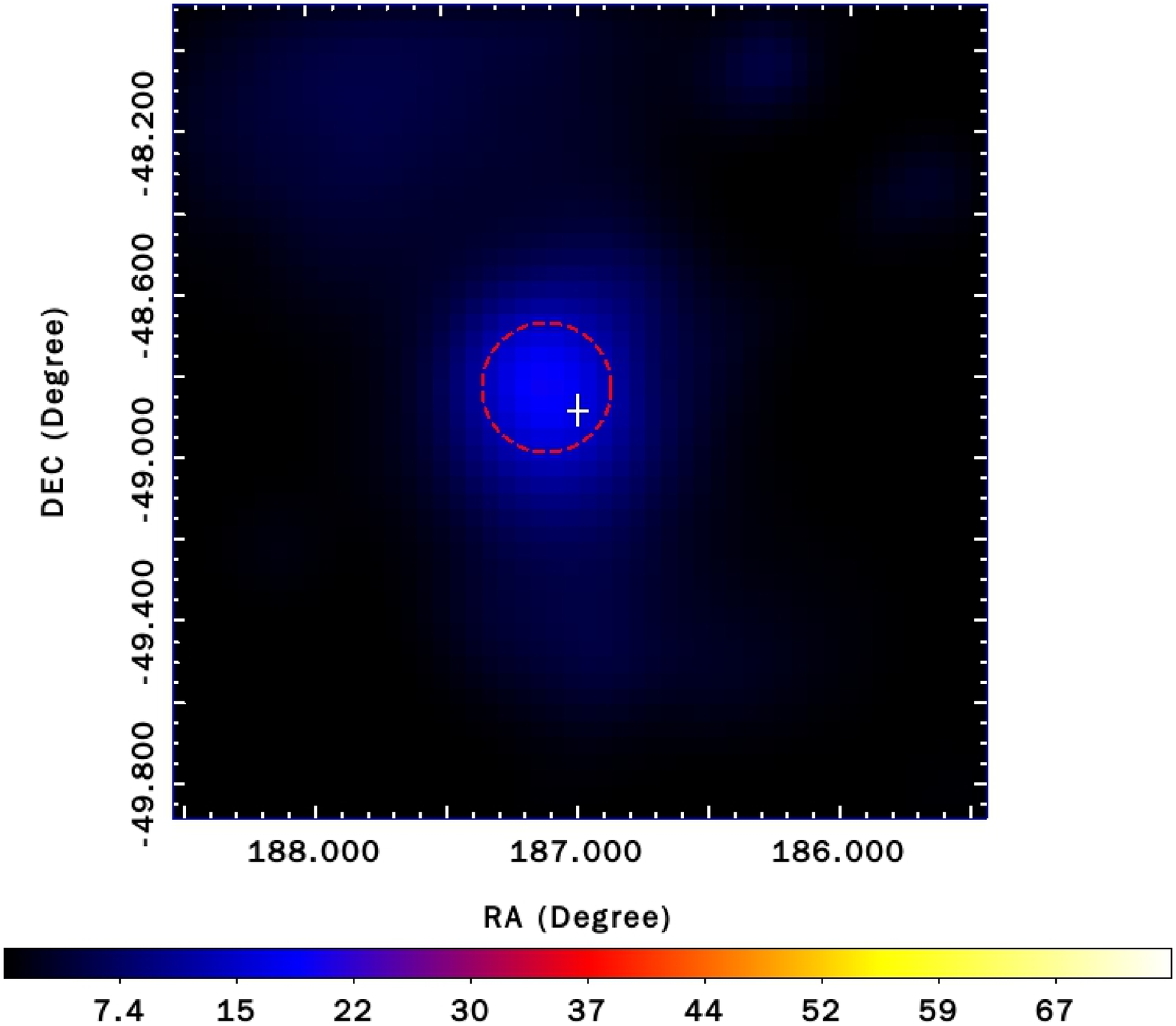}{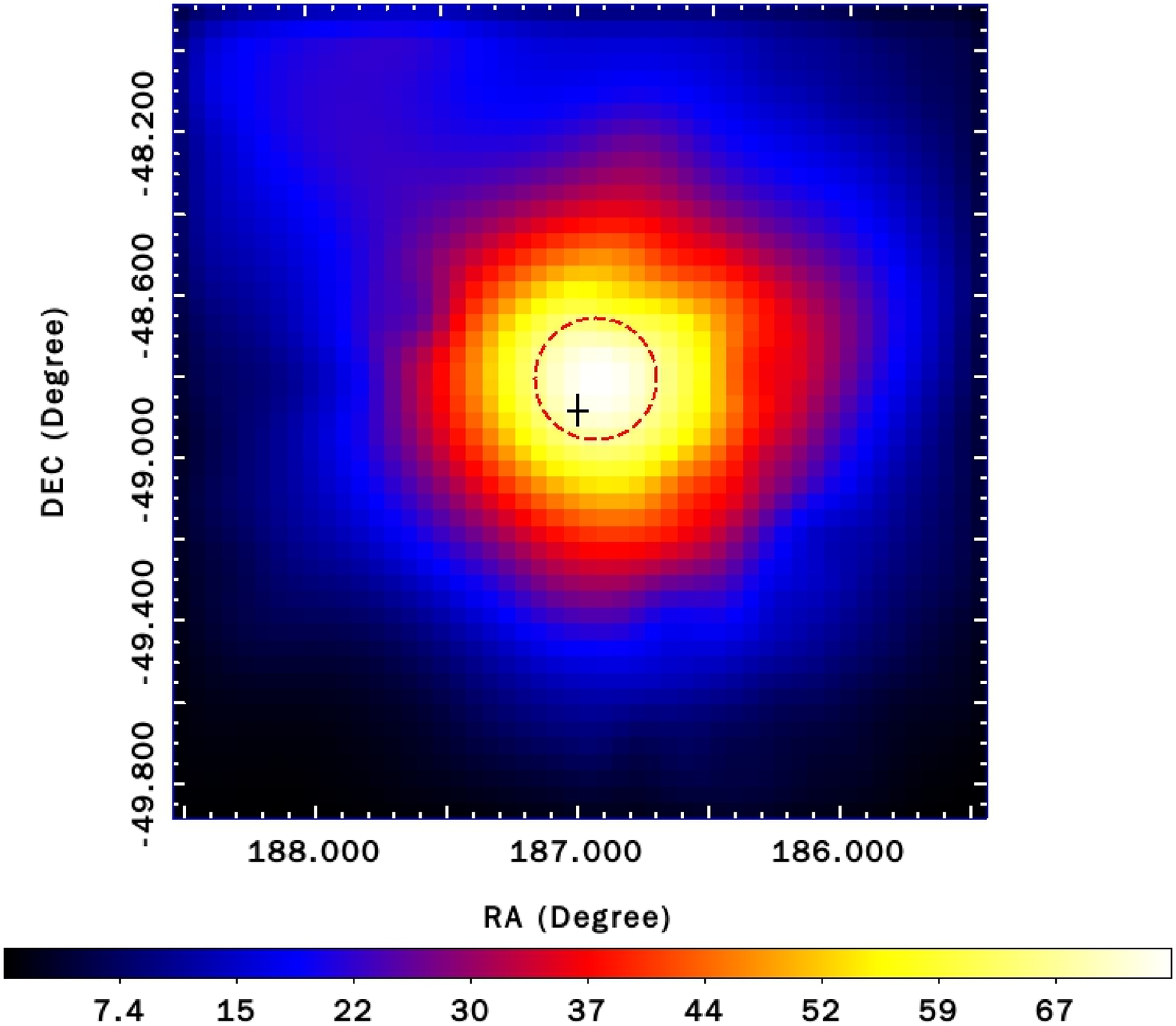}
\caption{0.3$-$300 GeV TS maps of a $\mathrm{2^{o}\times2^{o}}$ 
region centered at the position of XSS J12270$-$4859 for 
Phase I ({\it left panel}) and Phase II ({\it right panel}). 
The image scale of the maps is 0.04\arcdeg\ pixel$^{-1}$. 
The white ({\it left panel}) and dark ({\it right panel}) crosses mark 
the 2MASS position of XSS J12270$-$4859, 
and the dashed red circles indicate 
the 2$\sigma$ error circles of the best-fit positions of 
2FGL J1227.7$-$4853 during the two phases.}
\label{fig:tsmap-phase}
\end{figure}

We made two $>$0.3 GeV TS maps over   
the phase ranges of 0.1--0.4 (named Phase I) and 
0.6--0.9 (named Phase II),
which approximately are the bottom and peak, respectively, of the orbital
modulation (Figure~\ref{fig:timing}). 
The obtained TS maps are shown in Figure~\ref{fig:tsmap-phase}. 
The source during Phase II 
is more significantly detected than during Phase I, as the TS values 
are $\simeq$76 and $\simeq$20, respectively. 
We also ran \textit{gtfindsrc} to determine the positions of 
the $\gamma$-ray emission during the two phases,
and found that they are 
consistent with the position of XSS J12270$-$4859 within 2$\sigma$ error 
circles. 
The analysis confirms the detection of orbital modulation 
from the photon folding.

Spectra during Phase I and Phase II were also
obtained, but due to limited numbers of photons, the uncertainties on the 
flux data points are too large to allow any further detailed analysis.

\section{Discussion}
\label{sec:disc}

From our analysis of the \textit{Fermi} data of 2FGL J1227.7$-$4853, we have 
detected a flux decrease during the state transition
of XSS J12270$-$4859, which occurred around 2012 Nov.--Dec..
We have also detected orbital modulation, although
weakly, only in the data after the transition, consistent with
the result from the X-ray observation \citep{bog+14}. Both detections are 
strong evidence for the assocation between the two sources. In addition, 
the exponentially cutoff power-law model, which is a typical spectrum
for pulsar emission at the \fermi\ LAT \gr\ energy range \citep{1fpsr}, 
is preferred to
describe the source's emission during both before and after the transition 
time periods. This result supports the association as well. Given these,
our analysis has confirmed the previous identification that 
2FGL J1227.7$-$4853 is the \gr\ counterpart to XSS J12270$-$4859 
\citep{hil+11,mar+13}.

From extensive observations, particularly at \gr\ band, of the state transition 
of PSR J1023+0038 that occurred in 2013 June, 
it has been learned that in
the state of having an accretion disk, \gr\ emission is brighter than
that in the disk-free state: there was an order of magnitude flux increase 
in J1023+0038 accompanying the state transition \citep{sta+14,tak+14}. 
What we have detected in XSS J12270$-$4859 is the opposite
to that of the J1023+0038 case in 2013, while the flux change is smaller,
as the flux in the latter state is approximately two-times lower than 
that in the former.  It is very likely that \gr\ emission in the disk-free
state originates from the magnetosphere of the pulsar, while the brighter 
emission in the accretion state has been proposed to be due to inverse 
Compton (IC) scattering of a cold pulsar wind off
the optical/infrared photons from the accretion disk \citep{tak+14} or
self-synchrotron Compton processes at the magnetospheric region of
a propellering neutron star \citep{ptl14}.
Similar to that in J1023+0038, $\Gamma=1.8\pm 0.2$ and $E_c=2.3\pm0.9$\,GeV 
to 1.4$\pm$0.6 and 0.7$\pm$0.4\,GeV from the accretion state to disk-free
state, spectral changes were also detected in XSS J12270$-$4859, 
as $\Gamma\simeq 2.13$ and $E_c\simeq 6$\,GeV in the former changed to
$\simeq$1.8 and $\simeq$2\,GeV in the latter.
Although the large
uncertainties do not allow us to draw a clear conclusion, the measurements
are possible evidence for that the exact same physical processes 
occurred in XSS J12270$-$4859.

Probably because J12270$-$4859 is approximately 3 times brighter
than J1023+0038 in the disk-free state \citep{tam+10}, we have likely
detected its orbital modulation. Such \gr\ modulation, which was seen 
marginally in the black widow binary PSR B1957+20 (and the candidate 
MSP binary 2FGL J0523.3$-$2530; \citealt{wu+12}; \citealt{xwn14}), 
has been suggested to arise due to the view angle of the intrabinary 
interaction region \citep{wu+12,bed14}. However the modulation, approximately
aligned with that at X-rays \citep{bog+14}, has the brightness peak in 
the orbital phase of 0.5--1.0 (around the superior conjunction 
when the companion is behind the neutron star). This orbital variation is 
different from that
in B1957+20, as its brightness peak is at the opposite phase 
region. In the latter case the inverse Compton processes, 
which produces extra \gr\ emission around the inferior conjunction phase, 
has been suggested to be viewed as a head-on 
collision between the pulsar wind and the soft photons from the pulsar
or the companion \citep{wu+12,bed14}. For XSS~J12270$-$4859, we 
suspect its modulation may arise because of the occultation
of the photon emitting region by the companion, which well 
explains the X-ray modulation in PSR J1023+0038 (see \citealt{bog+11}
for details). We note that the inclination angle of the binary was
estimated to be
45\arcdeg--65\arcdeg \citep{mar+14}, also similar to that of PSR J1023+0038 
\citep{wan+09}.
Unfortunately, the photon counts were too low to allow a 
comparison of the phase-resolved spectra (obtained in the bright and faint
phase ranges), which might help identify the cause of the modulation.
If it were due to the occultation, no spectral changes would be expected.

%The $\gamma$-ray orbital modulation was observed to be due to the excess emission at several GeV energies, which was suggested to originated from inverse Compton (IC) scattering of the thermal radiation of the companion star by the unshocked `cold' ultrarelativistic pulsar wind \citep{wu+12}, or by the mixed, turbulent wind composed of pulsar wind and stellar wind \citep{bed14}. The changes of the viewing angle to the intrabinary $\gamma$-ray producing region as the binary rotates led to the observed $\gamma$-ray modulation.
In addition to the flux change related to the state transition,
our data analysis shows that \gr\ emission from XSS J12270$-$4859 may not
be stable before the state change, possibly having a factor of 3 flux variations in its long-term
light curve. The spectrum comparison between the high 
(with TS greater than 35) and low (with TS lower than 20) states only indicates 
the possible presence of an extra component around 0.33\,GeV; otherwise
similar flux changes were across the whole energy range.
Such flux variations are not seen before in MSP binaries. Recently in
the candidate MSP binary 2FGL~J0523.3$-$2530, significant flux variations were
detected but that was caused by the presence of a 2--3\,GeV component in 
the high state \citep{xwn14}. Given that there are three radio sources 
and a possible galaxy cluster in
the source field, contamination from them can not be totally excluded.
If one of them is associated with an unidentified blazer, which would
have caused the flux changes across the \textit{Fermi} energy range due to
its variability (e.g., \citealt{wil+14}), the flux variations would instead
indicate the contamination. In order to determine this possibility by detecting
any related flux changes at radio or X-ray energies, multi-wavelength
observations can be carried out when 
2FGL J1227.7$-$4853 shows significant brightening again.

\acknowledgements
This research was supported by supported by Shanghai Natural Science 
Foundation for Youth (13ZR1464400), the National Natural Science Foundation
of China (11373055), and the Strategic Priority Research Program
``The Emergence of Cosmological Structures" of the Chinese Academy
of Sciences (Grant No. XDB09000000). Z.W. is a Research Fellow of the
One-Hundred-Talents project of Chinese Academy of Sciences.

\bibliographystyle{apj}

\clearpage

\clearpage

\clearpage

\clearpage

\clearpage

\clearpage
\begin{deluxetable}{llcccc}
\tablecaption{Binned likelihood analysis results for 2FGL J1227.7$-$4853.}
\tablewidth{0pt}
\startdata
\hline
\hline
~ & Spectral model & Flux/10$^{-9}$ & $\Gamma$ & E$_{c}$ & TS \\
~ &  & (photon cm$^{-2}$ s$^{-1}$) &  & (GeV) &   \\
\hline
Total data & Power law & 30 $\pm$ 1 & 2.41 $\pm$ 0.03 & \nodata & 1446 \\
~ & Power law with cutoff & 28 $\pm$ 1 & 2.11 $\pm$ 0.08 & 6 $\pm$ 2 & 1466 \\
\hline
Before state change & Power law & 33 $\pm$ 1 & 2.42 $\pm$ 0.04 & \nodata & 1247 \\
~ & Power law with cutoff & 32 $\pm$ 2 & 2.13 $\pm$ 0.08 & 6 $\pm$ 2 & 1262 \\
\hline
After state change & Power law & 20 $\pm$ 2 & 2.42 $\pm$ 0.09 & \nodata & 183 \\
~ & Power law with cutoff & 17 $\pm$ 2 & 1.8 $\pm$ 0.3 & 2 $\pm$ 1 & 192 \\
\enddata
\label{tab:likelihood}
\end{deluxetable}

\clearpage
\begin{deluxetable}{lcccc}
\tablecaption{Flux measurements of 2FGL J1227.7$-$4853.}
\tablewidth{0pt}
\startdata
\hline
\hline
E & F$\mathrm{_{low}}$/10$^{-12}$ & F$\mathrm{_{high}}$/10$^{-12}$ & F$\mathrm{_{before}}$/10$^{-12}$ & F$\mathrm{_{after}}$/10$^{-12}$\\
(GeV) & (erg cm$^{-2}$ s$^{-1}$) & (erg cm$^{-2}$ s$^{-1}$) & (erg cm$^{-2}$ s$^{-1}$) & (erg cm$^{-2}$ s$^{-1}$)\\
\hline
0.15 & 13 $\pm$ 2 & 18 $\pm$ 5 & 13 $\pm$ 2 & 6 $\pm$ 3 \\
0.33 & 7 $\pm$ 1 & 21 $\pm$ 2 & 11.9 $\pm$ 0.8 & 6 $\pm$ 1 \\
0.74 & 8 $\pm$ 1 & 13 $\pm$ 2 & 9.5 $\pm$ 0.6 & 6 $\pm$ 1 \\
1.65 & 6.3 $\pm$ 0.9 & 11 $\pm$ 2 & 7.7 $\pm$ 0.6 & 5.7 $\pm$ 0.9 \\
3.67 & 2.9 $\pm$ 0.9 & 6 $\pm$ 2 & 4.1 $\pm$ 0.6 & 1.9 $\pm$ 0.7 \\
8.17 & \nodata & 5 $\pm$ 2 & 2.2 $\pm$ 0.6 & 0.8 $\pm$ 0.6 \\
18.20 & \nodata & 2 $\pm$ 2 & 1.2 $\pm$ 0.7 & \nodata \\
\enddata
\tablecomments{Columns 2 and 3 list the energy flux (E$^{2}$ $\times$ dN/dE) in each energy bin during the low and high states before the state change, respectively. Column 4 and 5 list the energy flux (E$^{2}$ $\times$ dN/dE) in each energy bin during the observations before and after the state change, respectively.}
\label{tab:spectrum-point}
\end{deluxetable}

\end{document}